\documentclass{aastex7}

\begin{document}

\title{Dual Origins of Rapid Flare Ribbon Downflows in an X9-class Solar Flare}

\author[0000-0001-9726-0738]{Ryan J. French}
\affiliation{Laboratory for Atmospheric and Space Physics, University of Colorado Boulder, 3665 Discovery Drive, Boulder, CO 80303, USA}
\email{ryan.french@lasp.colorado.edu}

\author[0000-0002-6368-939X]{William H. Ashfield IV}
\affiliation{Southwest Research Institute, 1301 Walnut St, Suite 400, Boulder, CO 80302, USA}
\email{}

\author[0000-0002-3229-1848]{Cole A. Tamburri}
\affiliation{National Solar Observatory, 3665 Discovery Drive, Boulder, CO 80303, USA}
\affiliation{Laboratory for Atmospheric and Space Physics, University of Colorado Boulder, 3665 Discovery Drive, Boulder, CO 80303, USA}
\affiliation{Department of Astrophysical and Planetary Sciences, University of Colorado Boulder, 2000 Colorado Ave, Boulder, CO 80305, USA}
\email{}

\author[0000-0001-8975-7605]{Maria D. Kazachenko}
\affiliation{National Solar Observatory, 3665 Discovery Drive, Boulder, CO 80303, USA}
\affiliation{Department of Astrophysical and Planetary Sciences, University of Colorado Boulder, 2000 Colorado Ave, Boulder, CO 80305, USA}
\affiliation{Laboratory for Atmospheric and Space Physics, University of Colorado Boulder, 3665 Discovery Drive, Boulder, CO 80303, USA}
\email{}

\author[0000-0002-1196-4046]{Marie Dominique}
\affiliation{Solar-Terrestrial Centre of Excellence—SIDC, Royal Observatory of Belgium, 3 Avenue Circulaire, B-1180 Uccle, Belgium}
\email{}

\author[0000-0003-1597-0184]{Marcel Corchado Albelo}
\affiliation{National Solar Observatory, 3665 Discovery Drive, Boulder, CO 80303, USA}
\affiliation{Laboratory for Atmospheric and Space Physics, University of Colorado Boulder, 3665 Discovery Drive, Boulder, CO 80303, USA}
\affiliation{Department of Astrophysical and Planetary Sciences, University of Colorado Boulder, 2000 Colorado Ave, Boulder, CO 80305, USA}
\email{}

\author[0000-0001-8702-8273]{Amir Caspi}
\affiliation{Southwest Research Institute, 1301 Walnut St, Suite 400, Boulder, CO 80302, USA}
\email{}

\begin{abstract}

We detect rapid downflows of 150--217~km/s in IRIS \ion{Si}{4} 1402.77~nm measurements of an X9-class solar flare on 2024 October 3rd. The fast redshift values persist for over 15~minutes from flare onset, and can be split into two distinct stages of behavior, suggesting that multiple mechanisms are responsible for the downwards acceleration of flare ribbon plasma. The first stage of rapid downflows are synchronized with peaks in emission from the Advanced Space-based Solar Observatory Hard X-ray Imager (ASO-S/HXI) and Large Yield Radiometer (LYRA) Lyman-$\alpha$ measurements, indicative that the chromospheric downflows (with a maximum redshift of 176~km/s) result from chromospheric condensations associated with impulsive energy release in the solar flare. Later in the event, strong \ion{Si}{4} flare ribbon downflows persist (to a maximum value of 217~km/s), despite the magnetic flux rate falling to zero, and high-energy HXR and Lyman-$\alpha$ measurements returning to background levels. This is reflective of downflows in the flare ribbon footpoints of flare-induced coronal rain.
Hard X-ray spectral analysis supports this scenario, revealing strong non-thermal emission during the initial downflow stage, falling near background levels by the second stage.
Despite these distinct and contrasting stages of ribbon behavior, \ion{Si}{4} Doppler velocities exhibit quasi-periodic pulsations with a constant $\sim$50~s period across the 15-minute flare evolution (independently of loop length). We deduce that these pulsations are likely caused by MHD oscillations in the magnetic arcade. Finally, we utilize machine learning K-means clustering methods to quantify line profile variations during the stages of rapid downflows. 

\end{abstract}

\accepted{to ApJ, Nov 2025}


\section{Introduction} \label{sec:intro}

Solar flares are the conversion of magnetic free energy into the acceleration of particles, plasma heating, and light across the electromagnetic spectrum. As a result of the energy release from the solar flare in the corona, the lower-altitude chromospheric and transition region layers experience subsequent rapid heating. The resulting increase in chromospheric emission resulting from the plasma heating creates structures called flare ribbons, which are magnetically connected to the flare release site in the corona.
The exact mechanisms causing the heating of these lower layers are still up for debate. In the original standard “CSHKP” model of solar flares \citep{Carmichael,Sturrock,Hirayama,Kopp}, energy deposition in the chromospheric flare ribbons come from the collision of non-thermal particles accelerated from the coronal reconnection site. Other works highlight the roles of further energy deposition processes in the chromosphere, including Alfve\'n waves originating from the flaring corona  \citep[e.g.,][]{Fletcher2008,Reep2016}, or thermal conduction caused by the temperature difference between the two regions \citep[e.g.,][]{Antiochos1978, Cheng1983}. 

In the non-thermal particle (beam-heating) scenario, energy deposition in the chromosphere results in strong non-thermal hard X-ray (HXR) sources, as a result of Bremsstrahlung emission from accelerated electrons \citep{brown1971,aschwanden2002,holman2011}. Non-thermal HXR emission in flares has also been found to exhibit similar behaviors in space and time to Lyman-$\alpha$ emission in the chromospheric flare ribbons, suggesting that Lyman-$\alpha$ is also a signature of impulsive non-thermal energy deposition in the chromosphere \citep{Greatorex2023,Majury2025}. Flare HXR emission, particularly at lower energies, is also associated with thermal sources -- both the `bulk' 15--25~MK sources dominating soft X-ray emission, and a `super-hot' ($\gtrsim$30~MK) component observed in intense flares, thought to be associated with the reconnection processes \citep[e.g.][]{CaspiLin2010,Caspietal2014}. These sources are distinctly coronal rather than chromospheric, and with different temporal behaviors \citep[e.g.,][]{Caspietal2015}.

Following the rapid heating of the chromosphere, the plasma within this layer of the atmosphere subsequently expands in all directions; upwards, towards the corona (which we call chromospheric evaporation or sublimation), and downwards, towards the photosphere (called chromospheric condensation). 
In solar spectroscopy, the chromospheric upflows and downflows during solar flares manifest themselves as observed Doppler blue-shifts and red-shifts, respectively. Blue-shifts from chromospheric evaporation in flare ribbons are frequently observed in ultraviolet \citep[e.g.,][]{Battaglia2015,Graham2015}, extreme-ultraviolet \citep[see review by][]{Milligan2015}, and even hard X-rays \citep[e.g.,][]{Liu2006}. Redshifts from chromospheric condensations are also routinely detected by ground and space-based observations, in H-$\alpha$ observations \citep{Ichimoto1984,Fisher1985,Wuelser1987,Canfield1987,Wuelser1989}, and in measurements from the Interface Region Imaging Spectrograph \citep[IRIS;][]{DePontieu2014} in the \ion{Si}{4} and \ion{Mg}{2} lines \citep{Kerr2015,Graham2015,Lorincik22,Kerr2022,Xu2023}. Typical IRIS \ion{Si}{4} and \ion{Mg}{2} redshift velocities are found in the range of $\sim$10--100~km/s, with the fastest \ion{Si}{4} downflows of 160~km/s reported by \citet{Xu2023} during an X1.3-class solar flare. Chromospheric condensations have previously been linked to reconnection-driven energy release and heating in solar flares \citep{Fisher1989,Ashfield2022}.

Doppler velocities in flare ribbons often exhibit quasi-periodic pulsations (QPPs). QPPs are observed in many solar flare observables, including ribbon Doppler velocity \citep[e.g.,][]{Brannon2015,Jeffrey2018,Lorincik22}, HXRs \citep[e.g.,][]{Parks1969,Chiu1970}, soft X-rays \citep{Simoes2015,hayes2019,Hayes2020}, and radio waves \citep[e.g.,][]{Kupriyanova2016,Carley2019}. QPPs have also been detected in flare ribbon intensity variations parallel and perpendicular to flare ribbon propagation \citep[][respectively]{Naus2022,French2021}, and in the reconnection flux traced by the flare ribbons \citep{Corchado2024}. 
Proposed mechanisms for the generation of QPPs generally fall on a spectrum between processes produced by bursty quasi-periodic magnetic reconnection, and those triggered by MHD oscillations. Reviews of QPPs and their many possible mechanisms are provided by \citet{nakariakov2009,vandoorsselare2016,McLaughlin2018,Kupriyanova2020,Zimovets2021}.

Flare-driven coronal rain can also produce spectroscopic signatures in flare ribbons. Coronal rain is apparent plasma downflows within coronal loops, formed by coronal condensation \citep{Parker1953}. 
Flare-driven coronal rain is one of the two primary types of coronal rain, with the second occurring in quiescent loops and prominences \citep{Antolin2020}. Spectroscopic observations of flare-induced coronal rain footpoints in the ribbon are rarer than measurements of the more impulsive chromospheric condensations. \citet{Pietrow2024} measured redshift values in H-$\alpha$ and \ion{Ca}{2} 854.2~nm emission within a flare-ribbon coronal rain footpoint (for an X13.3-class event), but found values exceeding the 70~km/s limit of the spectral window used.
\citet{Song2025} examined similar footpoint features for an X1.6 class event, and resolved supersonic redshift velocities of 137~km/s in the same H-$\alpha$ and \ion{Ca}{2} 854.2~nm lines. Flare ribbon sub-structure, as a result of flare-induced coronal rain, has been found to exist on scales of at most 80--200~km \citep{Jing2016} or even smaller \citep[$\sim$50~km;][]{Tamburri2025}.

In this study, we present measurements of rapid \ion{Si}{4} downflows from the 2024 October 3rd X-class solar flare, observed by the IRIS instrument. We investigate two distinct stages of strong red-shifts and their relationship with HXR and Lyman-$\alpha$ light curves, and HXR spectra. Furthermore, we analyze the periodicity of QPPs in the \ion{Si}{4} Doppler velocities and discuss atypical line profiles detected during the event. Combining the above, we build a case for two distinct stages of flare ribbon behavior, caused by impulsive energy release and flare-induced coronal rain, respectively.

\section{Observations} \label{sec:observations}

\begin{figure*}
\centering
\includegraphics[width=18cm]{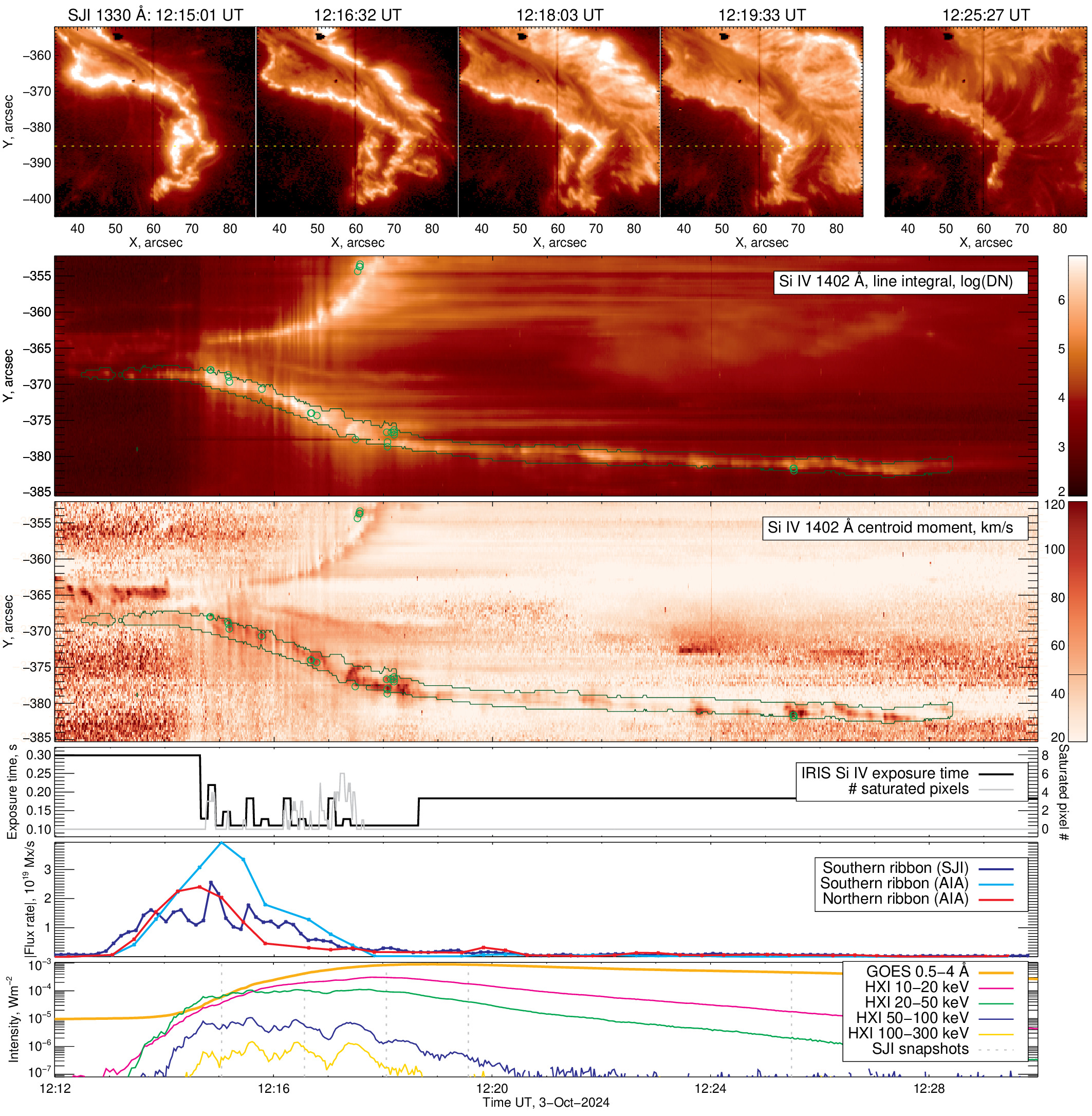}
\caption{
Overview of the X9.0 flare observed on 2024 October 3rd: Top row: SJI 1330~{\AA} snapshots (full FOV) of the flare ribbon evolution. The first four snapshots are sequentially 90~seconds apart during the impulsive phase, and final snapshot later in the gradual phase. The vertical black line is the location of the IRIS slit. The horizontal dashed line marks the lower extent of the FOV of subsequent panels.
Second row: Sit-and-stare \ion{Si}{4} 1402.77~{\AA} intensity map (along a sub-range of the slit), computed by line integration. The dark green line outlines the southern flare ribbon mask. Green circles highlight high redshift pixels examined in Figure~\ref{fig:spectra}.
Third row: Same as above, showing \ion{Si}{4} 1402.77~{\AA} Doppler velocity determined from line moment centroid.
Fourth row: Progression of IRIS \ion{Si}{4} 1402.77~{\AA} exposure time throughout the raster, alongside the number of saturated pixels per time-step.
Fifth row: Reconnection flux rate (magnitude) for the southern (AIA 1600~{\AA} and SJI 1330~{\AA}) and northern (AIA 1600~{\AA}) flare ribbons).
Bottom row: Time-series of GOES 0.5--4~{\AA} soft X-rays, and ASO-S HXI 10--20, 20--50, 50--100, 100--300~keV hard X-rays. ASO-S HXI measurements are multiplied by 1e--8 for comparison to GOES. Vertical dashed gray lines mark the location of SJI snapshots in the top row.
}
\label{fig:IRIS_context}
\end{figure*}

On 2024 October 3rd, the Sun produced an X9.0 solar flare from active region AR~13842 (close to the central meridian in the southern hemisphere), peaking at 12:18~UT. The flare produced two distinct flare ribbons, separating in the southeast-northwest direction.
The flare was observed by IRIS with a high-cadence flare program (OBS~4204700237), capturing sit-and-stare spectra of the IRIS small line list with a high cadence of 0.8~s. The IRIS 1330~{\AA} slit-jaw-imager (SJI) observed with a slower 8~s cadence over a $58{\arcsec} \times 62{\arcsec}$ field-of-view (FOV). IRIS observed with this program from 07:21:27--13:21:46~UT, covering the full flare duration, alongside several hours of pre-flare activity. 

The top row of Figure~\ref{fig:IRIS_context} shows five IRIS SJI 1330~{\AA} snapshots. The first four frames are separated by 90~s, showing flare ribbon evolution during the flare impulsive phase. 
The fifth frame shows a snapshot later in the flare, with decaying ribbons and flare loops visible. The main sections of both the northern and southern ribbons are within the SJI FOV during the impulsive phase, with the northern ribbon exiting the FOV a few minutes into the flare evolution.
The vertical black line in the center of the SJI marks the slit location. Fortuitously, the slit is centered on both flare ribbons, intersecting both at flare onset, and losing the northern ribbon around 12:18~UT. Sections of the southern ribbon remain under the slit until they fully decay. In this study, we focus on spectral observations of the \ion{Si}{4} 1402.77~{\AA} line. The line is \textit{typically} believed to behave under the optically thin coronal assumption, even during flaring conditions \citep{Brannon2015,Kerr2019b,Kerr2022,Ashfield2022}.  
Conversely, there are some studies suggesting that opacity may play a role in the formation of \ion{Si}{4} 1402.77~{\AA} during transient periods \citep[e.g.,][]{Zhou2022,Kerr2023}.
In addition to the \ion{Si}{4} 1402.77~{\AA} line, the high-cadence sit-and-stare raster also includes measurements of the \ion{C}{2} 1336~{\AA} and \ion{Mg}{2}~k 2796~{\AA} lines, but we do not investigate these in this work.

The second row of Figure~\ref{fig:IRIS_context} shows the sit-and-stare raster of logarithmic integrated line intensity across the full \ion{Si}{4} 1402.77~{\AA} line. Because of the complex line profiles, we do not fit Gaussian profiles at this stage, instead measuring the total emission from the integrated area of the line. Similarly, the third row of Figure~\ref{fig:time_series} shows the Doppler velocity \textit{moment} of the total line emission -- the bulk velocity shift of the emission line weighted-average centroid. This Doppler velocity moment measures an average Doppler velocity over the spectral profile, not accounting for multiple higher-velocity red- or blue-shifted components that may be convolved into a single emission line. A closer investigation (and precise fitting) of individual spectral lines is analyzed in Section~\ref{sec:redshift}. The y-axes of the intensity and velocity sit-and-stare raster panels in Figure~\ref{fig:IRIS_context} are cropped from the full slit range, with a lower extent marked by the horizontal dashed line in the top row of Figure~\ref{fig:IRIS_context}. The sit-and-stare rasters show the formation of both flare ribbons, before the northern ribbon exits the FOV. The dark green line shows a mask of the southern flare ribbon, calculated by including all pixels either above a background level (600~DN), or within $\pm$2~pixels of a polynomial curve fit to pixels over 40\% of the maximum intensity at each time step. These threshold ranges are chosen manually to capture an area reflective of the primary flare ribbon structure.  This mask is later used to generate light curves of intensity and Doppler velocity along the evolving flare ribbons, presented in Figure~\ref{fig:time_series}. We also analyze spectral profiles within this same southern mask region.


The brightest pixels experience saturation in the \ion{Si}{4} 1402.77~{\AA} line, despite IRIS's variable exposure rate during flares. In some pixels, saturation is still present at exposure times of 0.11~seconds -- the lowest achievable by the instrument. The panel in the fourth row of Figure~\ref{fig:IRIS_context} plots the change in exposure time (solid black line, left axis), alongside the number of saturated pixels at each time step (light gray line, right axis).

For an additional metric of energy release during the flare, we utilize photospheric magnetic field measurements from the Solar Dynamics Observatory's Helioseismic and Magnetic Imager \citep[HMI;][]{Schou2012} to track the estimated rate of magnetic reconnection flux with time throughout the flare. The magnitudes of these reconnection fluxes are presented in the fifth row of Figure~\ref{fig:IRIS_context}. This metric is obtained by measuring the HMI vertical magnetic flux enclosed by the sweeping flare ribbon region \citep{Forbes1984,Forbes2000,Fletcher2001,Qiu2004,Kazachenko2017,Corchado2024}, and taking the derivative of the reconnection flux to find the rate of magnetic reconnection flux. We complete this analysis for the ribbons observed in both SJI 1330~{\AA} images (for the southern ribbon within the FOV) and Solar Dynamics Observatory / Atmospheric Imaging Assembly \citep[SDO/AIA;][]{Lemen2012} 1600~{\AA} images (for both northern and southern ribbons, albeit at a lower spatial and temporal resolution than SJI observations).

The sixth row of Figure~\ref{fig:IRIS_context} shows simultaneous X-ray measurements of the solar flare. The orange curve presents GOES XRS 0.5--4~{\AA} soft X-rays, peaking at $9\times10^{-4}$~Wm$^{-2}$, classifying the flare at an X9 level. Subsequent light curves plot background-subtracted HXR emission measured by the Hard X-ray Imager \citep[HXI;][]{Zhang2019} onboard the Advanced Space-based Solar Observatory
\citep[ASO-S;][]{Gan2019}. The HXR time series are plotted in count rates normalized by $10^8$, in order to present the data on the same axis as GOES XRS emission. We plot HXI emission in energy bins of 10--20, 20--50, 50--100, and 100--300~keV. The lower y-axis limit is set at the noise level of the HXR emission. 

The event was also observed by the Large Yield Radiometer \citep[LYRA;][]{Hochedez2006,Dominique2013} on the 
Project for On-Board Autonomy \citep[PROBA-2;][]{Santandrea2013} spacecraft, in the Lyman-$\alpha$ 120--123~nm channel. Lyman-$\alpha$ observations from LYRA are presented in Figure~\ref{fig:time_series}.

\section{Analysis} \label{sec:analysis}
\subsection{Time Series Analysis}

\begin{figure*}
\centering
\includegraphics[width=18cm]{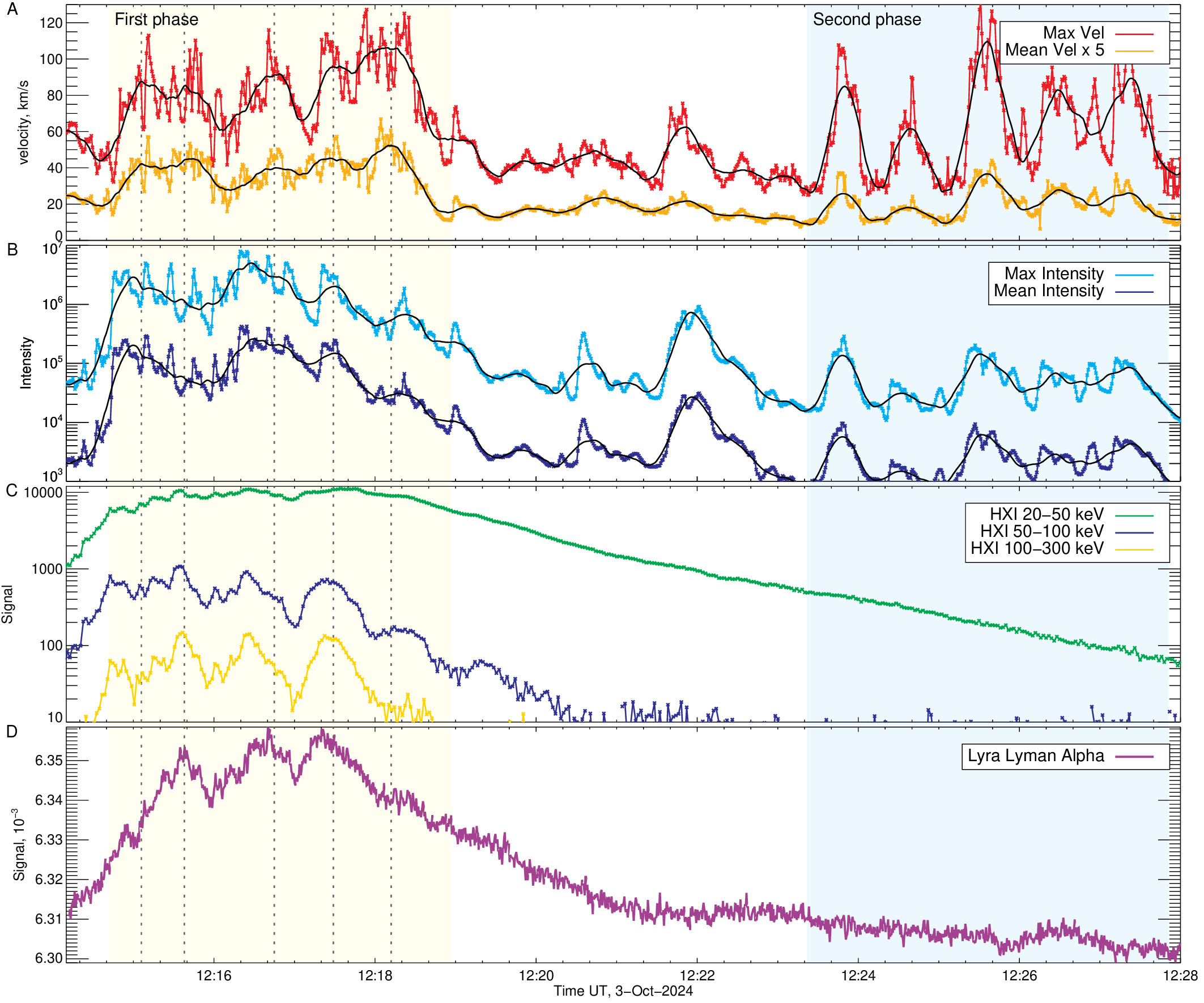}
\caption{A: Max and mean ($\times5$) \ion{Si}{4} 1402.77~{\AA} Doppler velocity across ribbon mask. Black line marks smoothed time series.
B: Max and mean \ion{Si}{4} 1402.77~{\AA} intensity across ribbon mask. Black line marks smoothed time series.
C: ASO-S HXI HXR time series.
D: LYRA Lyman-$\alpha$ time series.
 Vertical dashed lines mark the locations of local peaks in the smoothed maximum \ion{Si}{4} Doppler velocity (black line).
 We use yellow and blue background shading to highlight our two stages of rapid \ion{Si}{4} Doppler velocities -- the impulsive ($\sim$12:15--12:19~UT) and non-impulsive ($\sim$12:23--12:28~UT) phases.
}
\label{fig:time_series}
\end{figure*}

\begin{figure*}
\centering
\includegraphics[width=18cm]{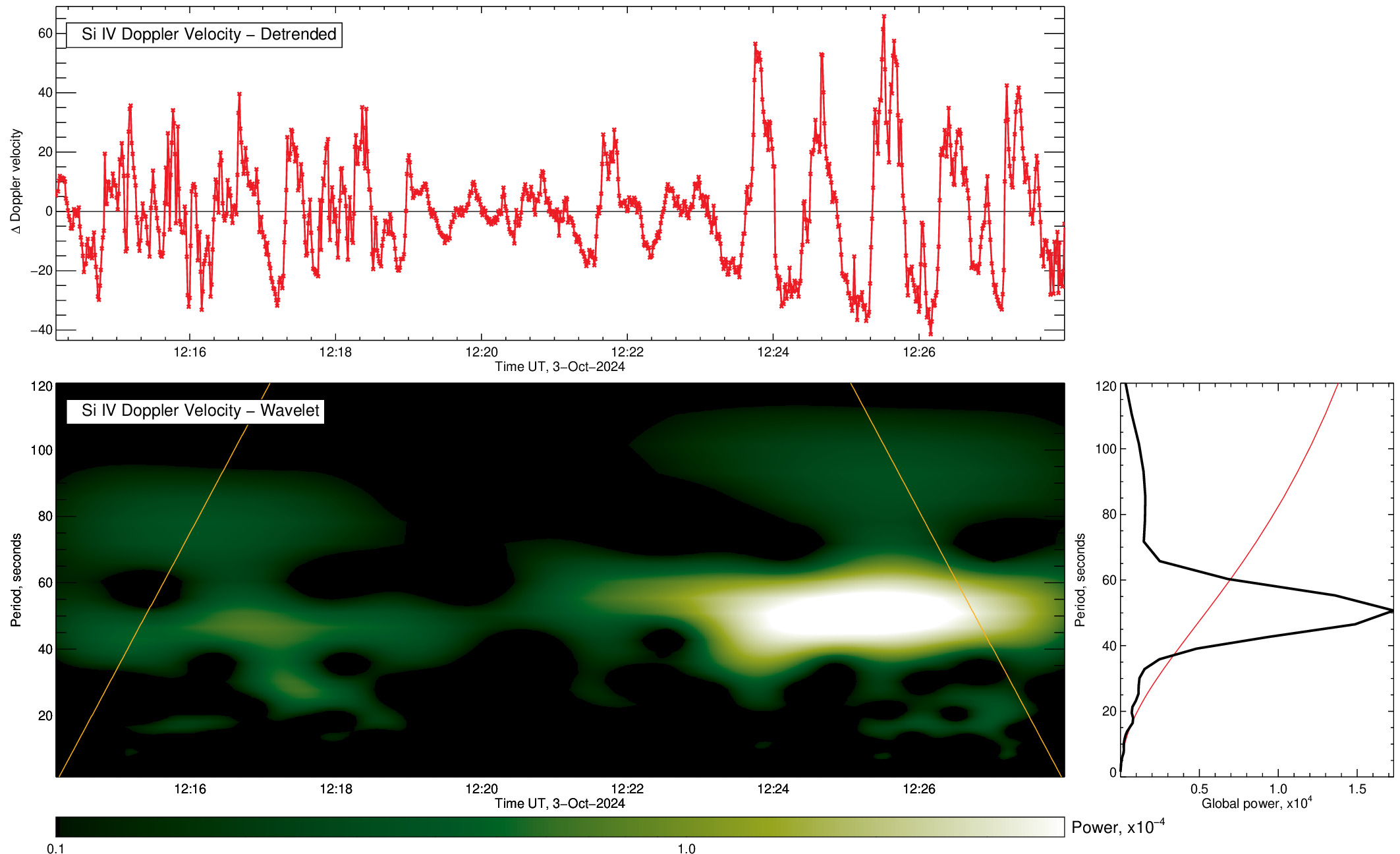}
\caption{Top: 120-second detrended \ion{Si}{4} Doppler velocity within the southern flare ribbon (detrended version of the red maximum velocity curve in Figure~\ref{fig:time_series}. Bottom-left: Wavelet power spectrum of time-series plotted in top panel (on a log color scale). Orange lines mark the cone of influence. Bottom-right: Global wavelet power for the same dataset. Red line marks the 95\% confidence threshold.
}
\label{fig:wavelet}
\end{figure*}  

We take a time-series slice through the sit-and-stare intensity and Doppler velocity rasters in Figure~\ref{fig:IRIS_context}.
Because the southern ribbon is visible for a longer duration than the northern ribbon, and exhibits faster average Doppler velocities in Figure~\ref{fig:IRIS_context}, we focus here only on the time series of the southern ribbon evolution, during 12:14--12:28~UT. 
Figure~\ref{fig:time_series}A presents \ion{Si}{4} Doppler velocity moment variation along the flare ribbon mask plotted in Figure~\ref{fig:IRIS_context}, plotting both the maximum velocity at each time step along the masked sit-and-stare raster, and mean velocity within the mask at each time step, as calculated via moment analysis. 
Similarly, Figure~\ref{fig:time_series}B presents both the maximum and mean intensity along the mask at each time step. 
The colored curves in Figure~\ref{fig:time_series}A--B show the intensity and Doppler velocity at maximum cadence. The black curves plot a 45-s-smoothed moving average of the colored curves, to remove shorter-duration variations and better reveal the longer-duration evolution in the time series data.
Figure~\ref{fig:time_series}C takes a closer look at the higher-energy ASO-S HXR measurements also presented in Figure~\ref{fig:IRIS_context}, and panel D shows the LYRA Lyman-$\alpha$ emission. For the Lyman-$\alpha$ data, we used the calibrated data (level~2) from the backup unit (Unit~1) of LYRA, which is the least affected by degradation. We implement a 10$\times$ binning to improve the signal-to-noise of the data.


The (45~s) smoothed \ion{Si}{4} Doppler velocity moment and intensity curves in Figure~\ref{fig:time_series}A--B
reveal two distinct stages of high-amplitude QPP-like oscillations. The first stage of oscillations, which is more easily visible in the velocity time series than intensity, occurs around $\sim$12:15--12:19~UT (shaded yellow in Figure~\ref{fig:time_series}). This is within the impulsive phase of the flare, during the period of enhanced Lyman-$\alpha$ emission and HXR emission in the 50--100~keV and 100--300~keV channels that are typically associated with impulsive energy release. The HXR emission also displays oscillations of similar periods, with peaks close in time to the smoothed peaks in \ion{Si}{4} velocity and intensity. The three most prominent localized peaks are also present at the maximum of the Lyman-$\alpha$ light curve.

The second sequence of \ion{Si}{4} oscillations comes later in the time series, with clear peaks in the black smoothed-average curves around $\sim$12:23--12:28~UT (shaded blue in Figure~\ref{fig:time_series}). 
During these oscillations, peak amplitudes of the downflow velocities are comparable with those seen during the impulsive phase of the flare. At the largest peak, the maximum velocity moment swings from 30~km/s to 130~km/s and back, in periods under one~minute. Although the oscillations are clearer in the time series of maximum Doppler velocity and intensity, they are still clear in the time series of the mean values, too. 

During this later sequence of oscillations, the higher-energy HXR emission has already fallen below the background noise level. Similarly, there are no clear peaks present in Lyman-$\alpha$ emission. This is indicative that the earlier impulsive particle acceleration has now reduced significantly.
The reconnection rate magnitudes presented in the fifth row of Figure~\ref{fig:IRIS_context} are also in agreement with this scenario, finding reconnection rates have fallen to near-zero by 12:21~UT -- prior to the second stage of the \ion{Si}{4} oscillations. The low reconnection rate therefore adds further evidence that electron precipitation has reduced below observable levels. This is consistent with expectations from the lack of impulsive HXR or Lyman-$\alpha$ signatures, as reconnection rates evolve co-temporally with particle acceleration proxies from HXR observations \citep[see, e.g.,][]{Miklenic2006,qiu2010,Veronig2015,tamburri2024}.

We examine the period of the Doppler velocity oscillations using wavelet analysis \citep[Mortlet method;][]{torrence_compo} in Figure~\ref{fig:wavelet}. We detrend the maximum Doppler velocity timeseries from Figure~\ref{fig:time_series} with a Savitzky–Golay filter \citep{savgol_1964}, as per the methodology for analyzing QPPs laid out in \citet{Broomhall2019}. We detrend with a period of 120~s. The resulting detrended time series is presented in the top panel of Figure~\ref{fig:wavelet}, which is used as the input for the wavelet analysis. The bottom panels show the corresponding wavelet power spectrum and global wavelet spectrum. The global wavelet captures an average period of 50~s across the time series. Curiously, the 2D wavelet power spectrum reveals that the $\sim$50~s oscillation period is present throughout the time window, near-constant throughout both the distinct impulsive and non-impulsive (with and without near-corresponding HXR and Lyman-$\alpha$ emission) phases of the rapid Doppler velocity oscillations.

\subsection{X-ray Spectral Analysis}

\begin{figure*}
\centering
\includegraphics[width=18cm]{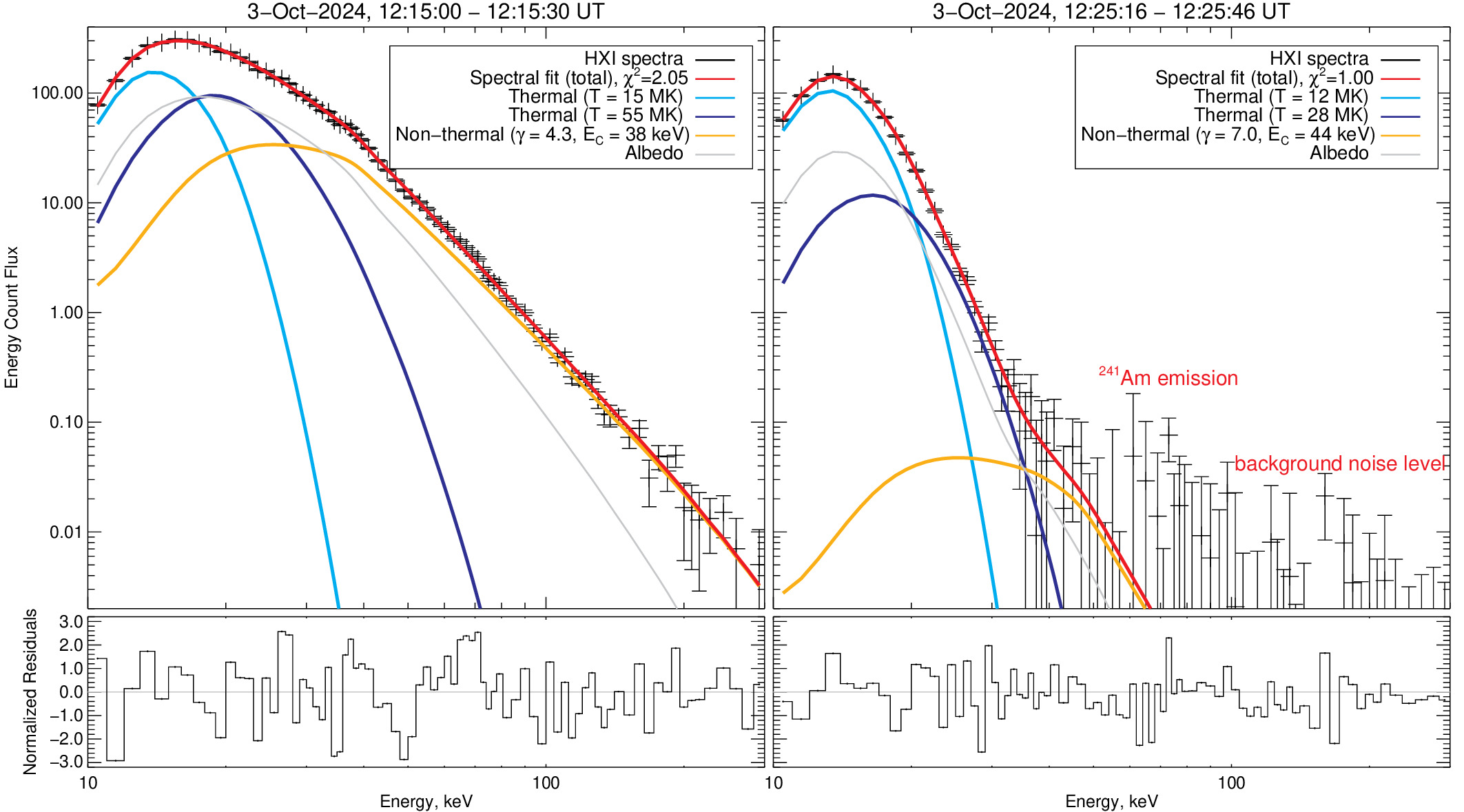}
\caption{
Top: Hard X-ray spectra (black markers) from ASO-S/HXI, within the time windows of 12:15:00--12:15:30~UT (left panel) and 12:25:16--12:25:46~UT (right panel). This time periods encapsulate part of our two stages of interest, with the former focusing on the peak of the HXR emission, and latter centered around the strongest \ion{Si}{4} downflow observed in the later stage. In the second panel, we label in the data the locations of residual background emission from $^{241}$Am (see text), and the residual background noise level.
The red curve in each panel presents the total spectral fit, consisting of two isothermal components and a non-thermal component (cyan, blue and orange, respectively), with key parameters as noted in the legend; a photospheric albedo correction (gray curve) is also included.
Bottom: The fit residuals (the differences between black markers and red curves) normalized by respective uncertainties in each energy bin.
}
\label{fig:hxi_spectra}
\end{figure*}

Figure~\ref{fig:time_series} reveals that 50--100~keV and 100--300~keV HXR emission has fallen to background levels by the second stage of rapid \ion{Si}{4} downflows, but that 20--50~keV HXR emission still persists. To determine whether this continued emission is evidence of ongoing particle acceleration (which could suggest alternative downflow origins beyond flare-induced coronal rain), we must examine the partition of thermal to non-thermal emission in the HXR spectra. Figure~\ref{fig:hxi_spectra} presents HXR spectra within two 30~s windows. The first panel, encompassing emission during 12:15:00--15:15:30~UT, is centered around the peak of impulsive HXR emission and coincides with the first stage of observed rapid \ion{Si}{4} downflows. The second panel shows HXR spectra during 12:25:16--12:25:46~UT, centered on the fastest \ion{Si}{4} downflow observed during the later period of interest. The black markers show the observed count flux data. We use HXI detector D94, with a thin aluminum entrance window, to maximize sensitivity \citep{Su2023}. To isolate the flare signal, we first subtract the cotemporal measurements from the nearest background (BKG) detector, D99, which does not observe the Sun but measures only the background X-ray emission in the instrument (e.g., due to particle-induced bremsstrahlung and/or activation of the materials surrounding the detector). We then also subtract a fixed pre-flare spectrum, taken as the average of these BKG-subtracted measurements during the pre-flare period 12:00:01--12:02:01~UT. Error bars shown are based on standard Poisson statistics, $\sigma(E) = \sqrt{(I(E)}$, where $I(E)$ are the total counts at energy $E$, with appropriate propagation of error from the background subtraction and normalization to flux units. In the later time interval (right panel), the solar HXR signal has dropped sufficiently to reveal residual background at energies above $\sim$50~keV -- while some counts remain, these are consistent with Poisson fluctuations around zero atop minor orbital background variation.\footnote{The small residual peak around $\sim$59~keV is due to minor differences in the activities of the $^{241}$Am calibration sources embedded in each detector \citep{Su2024, ZLi2025},
but these differences are negligible compared to the flare peak fluxes in the left panel and do not affect the fit at later times in the right panel.}

To determine the relative contributions of thermal and non-thermal emission to the spectra during these two time periods, we use the IDL Object Spectral Executive \citep[\texttt{OSPEX};][]{OSPEX} code, one of the standard tools forward-fitting of solar HXR data. After background subtraction as detailed above, we follow a similar fit procedure to that used by \citet{CaspiLin2010} for RHESSI flares, where our emission model includes two isothermal components to account for both the `bulk' hot and additional `super-hot' emission (typical of these intense flares), along with a power-law non-thermal component, and an `albedo' correction to account for energy-dependent photospheric X-ray Compton backscatter \citep{Kontar2006}. These components are forward-modeled through the instrument response, and their parameters optimized via iterative chi-squared minimization. The thermal components assume standard coronal abundances \citep{Feldman1992}, and the temperature ($T$) and volume emission measure are free fit parameters. The power-law normalization, spectral index ($\gamma$), and low-energy `rollover' value (corresponding to a low-energy cutoff value for the non-thermal electrons) are freely fit; the photon spectral index below the rollover is fixed at 2, following previous RHESSI works. The best-fit individual model components and summed model spectrum are shown as the colored curves in the respective panels in Figure~\ref{fig:hxi_spectra}, along with uncertainty-normalized fit residuals below the spectra. The residuals are well distributed around zero with no significant systematic deviations, and the reduced chi-square values are low, indicating acceptable fits.


During the flare HXR peak, in the first time interval, we find best-fit isothermal component temperatures of $\sim$15~MK and $\sim$55~MK, which are in-family with temperatures and scaling relationships observed in prior statistical studies of intense flares \citep[e.g.,][]{Caspietal2014, WarmuthMann2016}. As expected, we also find significant non-thermal emission, with a photon spectral index $\gamma \approx 4.3$ above the low rollover energy $E_C \approx 38$~keV, corresponding to non-thermal electron energy deposition of $\sim$6.9$\times 10^{28}$~erg/s into the chromosphere \citep{Linetal2001}. In the second interval of interest, ten minutes later (during the gradual phase), we find that temperatures have cooled to $\sim$12~MK and $\sim$28~MK, respectively. Notably, these two thermal components, and particularly the hotter one, account for $>$97\% all the flux in the 20--50~keV lightcurve in Figure~\ref{fig:time_series}. The fit non-thermal component, while non-zero, is over 300 times weaker than in the prior interval -- the spectral index has softened to $\gamma \approx 7.0$ above a low rollover energy $E_C \approx 44$~keV, corresponding to a much-reduced non-thermal electron energy deposition of only $\sim$1.9$\times 10^{26}$~erg/s. Analysis of HXI detectors D92 and D93, with thicker aluminum filters, yields qualitatively similar results -- we find somewhat lower non-thermal emission and somewhat higher (hotter) super-hot emission, but nonetheless the significant relative drop of non-thermal emission between the first and second time intervals remains comparable. We present the D94 results here since the much lower HXR flux in the second time interval yields significantly higher signal-to-noise in that detector.


Comparing the two spectral fits, we deduce that significant non-thermal emission is present during the period of initial rapid \ion{Si}{4} downflows (corresponding to strong 50--100~keV and 100--300~keV HXR emission in Figure~\ref{fig:time_series}). Contrastingly, during the second stage of rapid downflows, even though the \ion{Si}{4} downflows are of similar magnitude, the non-thermal emission has decreased by well over two orders of magnitude; almost all of the remaining HXR flux is attributable to thermal emission. While some weak non-thermal particle acceleration and energy deposition remains, it clearly cannot account for the same magnitude of \ion{Si}{4} Doppler velocities observed during this later time period.

\subsection{Ribbon Redshifts} \label{sec:redshift}

\begin{figure*}
\centering
\includegraphics[width=18cm]{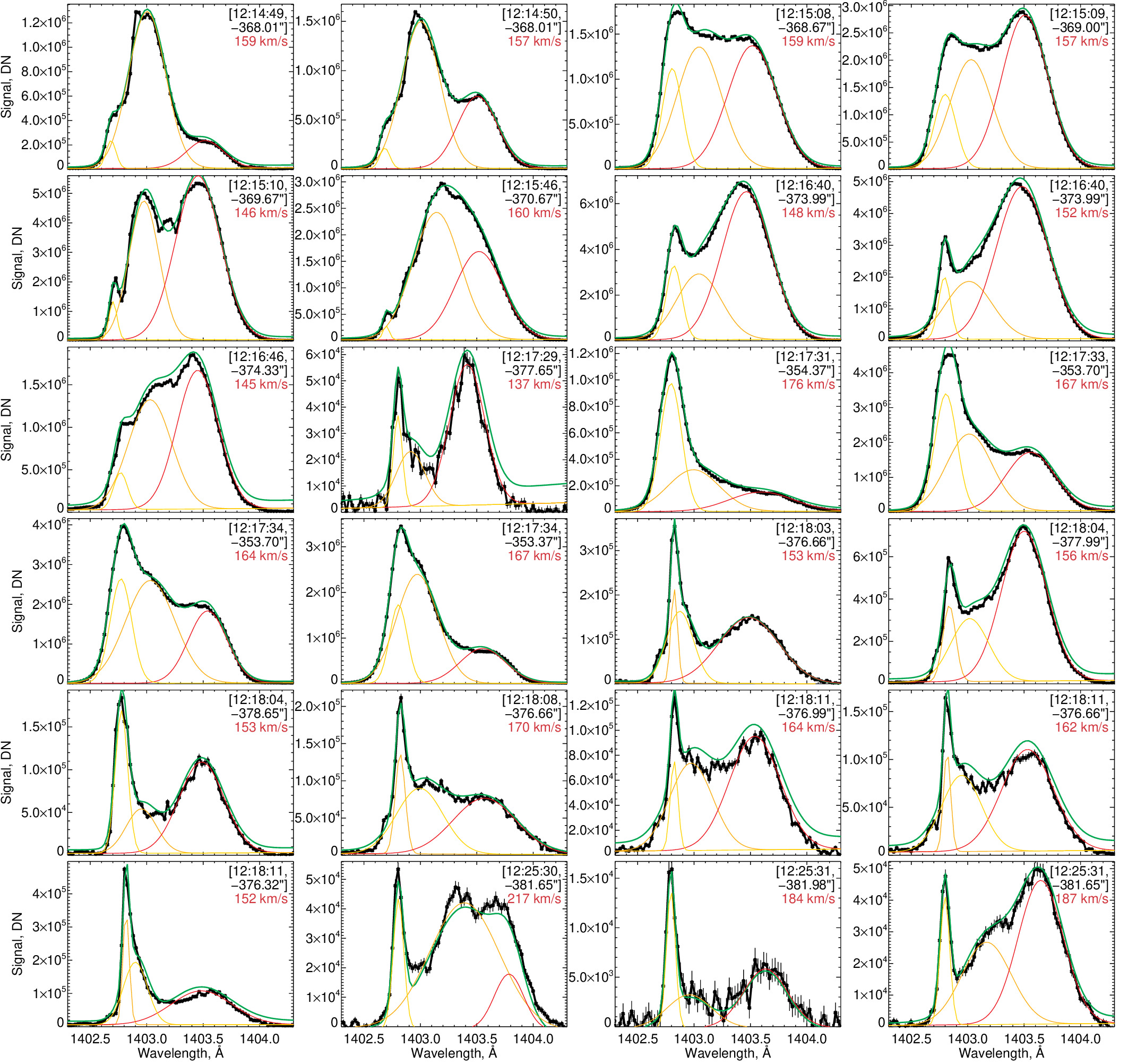}
\caption{\ion{Si}{4} 1402.77~{\AA} line profile and triple-Gaussian fit for a selection of the fastest red shift pixels across the sit-and-stare raster (plotted as green circles in Figure~\ref{fig:IRIS_context}). Black line plots the spectral data (with error bars), green the sum of the triple Gaussian fit, and yellow/orange/red the individual Gaussian fit profiles. Pixel position and velocity of the red-most Gaussian are noted in the top-right corner of each panel.
}
\label{fig:spectra}
\end{figure*}  

The velocity values presented in Figures~\ref{fig:IRIS_context} and~\ref{fig:time_series} are the Doppler moment velocity, corresponding to the mean wavelength shift across the full line profile. The decision to present velocities this way was chosen due to the complexity of the flare ribbon line profiles during the flare. To get a closer look at the true velocity components in the flare ribbons, we individually analyze spectra from a selection of the fastest velocity pixels. 

Figure~\ref{fig:spectra} presents the \ion{Si}{4} 1402.77~{\AA} spectra from a selection of high-velocity pixels, marked by green circles in Figure~\ref{fig:IRIS_context}. For each spectrum, we fit a triple-Gaussian profile. These Gaussians are overplotted in Figure~\ref{fig:spectra} in yellow, orange, and red -- ordered from least to most red-shifted. The presented pixels contain the strongest velocities within the most red-shifted Gaussian component of the fit (labeled in red text within each panel). For groups of directly adjacent pixels of fastest velocities, only the fastest was presented in this figure. The green line shows the total sum of the three Gaussian curves. These strongest-component pixels differ slightly from the pixels containing the fastest centroid-weighted Doppler velocities, which were plotted in Figures~\ref{fig:IRIS_context} and~\ref{fig:time_series}. 

A triple-Gaussian fit is often used to fit flare ribbon spectra \citep[e.g.,][]{Xu2023} to capture bulk plasma flows above and below a stationary layer, whilst the chromosphere is expanding (blueshift indicates evaporation, and redshift indicates condensations). Examining the pixel fits in Figure~\ref{fig:spectra}, we find multiple instances of peak redshifts of 150--170~km/s during the impulsive phase of the solar flare. Later in the flare evolution, we detect redshift values of 217~km/s, the fastest seen during the event. The line profiles throughout the flare evolution are complex, even unusual in some cases, varying significantly over small time periods or spatial separations. In many examples during the impulsive phase, the triple Gaussian is well resolved, with three bumps visible within the spectral profile. During the fast downflows seen later in the flare evolution, the three individual peaks are even clearer, with a larger separation between the blue-most and red-most Gaussian components.


We find that a majority of the pixels are accurately fit by a triple-Gaussian line profile. However, some \ion{Si}{4} spectra show complex line profiles, suggesting complexities in the line emission mechanism beyond that represented by the triple Gaussian profile, such as self-absorption or asymmetric line broadening processes. Under the assumption that \ion{Si}{4} 1402.77~{\AA} is optically thin in typical flaring temperatures, these features could be attributed to cool gas along the LOS of the \ion{Si}{4} emission, which has been used to explain \ion{Si}{4} profiles in UV bursts observed using IRIS\citep{Vissers2015}. Conversely, the complex \ion{Si}{4} profiles may be evidence for an optically thick regime \citep[e.g., as explored in][]{Kerr2019b,Zhou2022}. 

\subsection{Spectral Clustering} \label{sec:kmean}

\begin{figure*}
\centering
\includegraphics[width=18cm]{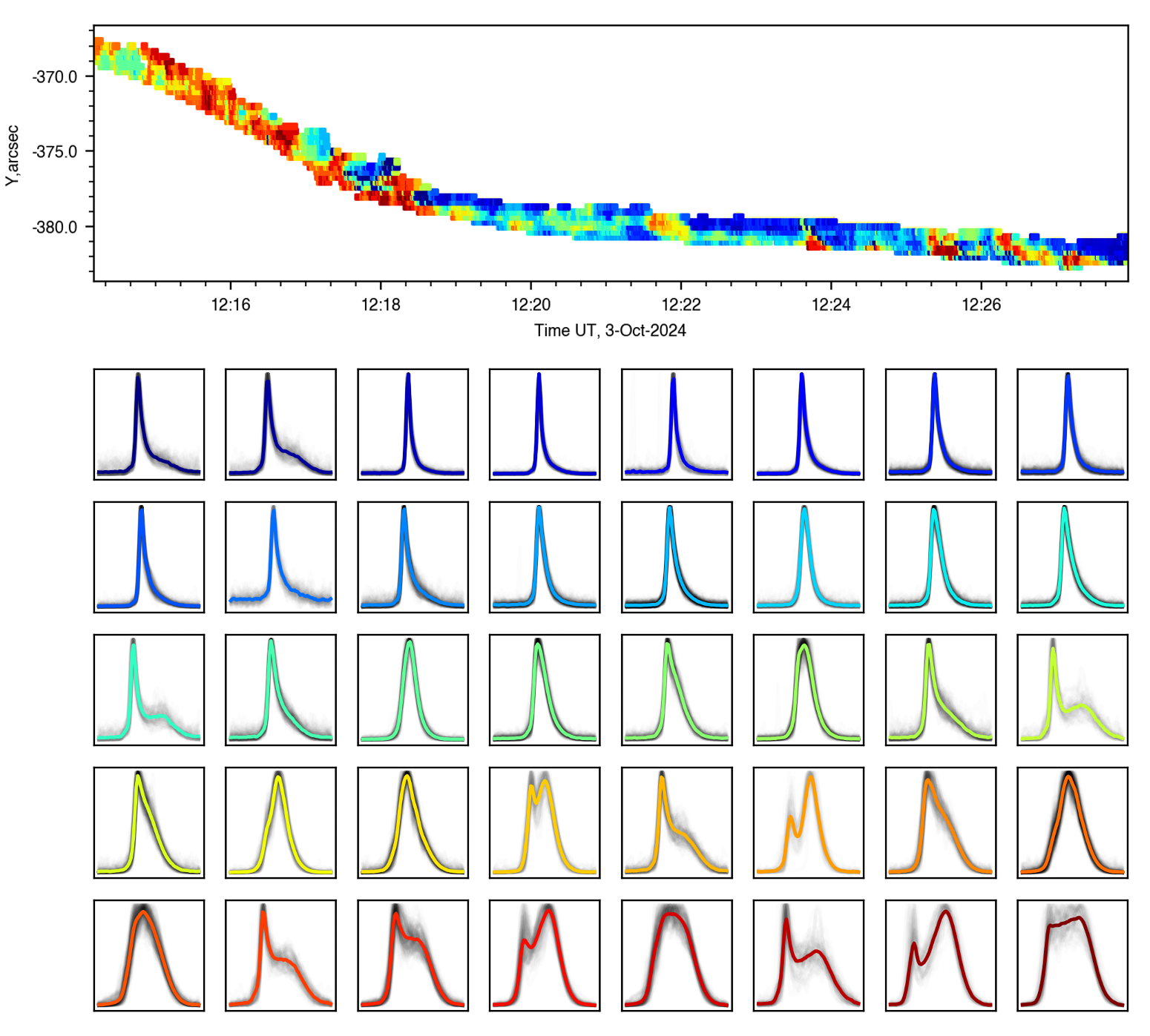}
\caption{Top: K-means clustering map of \ion{Si}{4} 1402.77~{\AA} spectra, marking the location of colored grouped spectra from panels below.
Bottom: 40~clustering groups of \ion{Si}{4} 1402.77~{\AA} spectra. Profiles are ordered by the full-width at 30\% of maximum, sorted from lowest to highest.
}
\label{fig:clustering}
\end{figure*}  

Figure~\ref{fig:spectra} offers a tantalizing insight into the diverse line profiles present during the solar flare. To get a better understanding of how these line profiles evolve in space and time, we conduct k-means clustering analysis on the ribbon spectra. The results of this analysis are presented in Figure~\ref{fig:clustering}.

We use the python \texttt{nltk} package\footnote{https://www.nltk.org/} to perform \textit{k}-means clustering \citep{bird2009} of \ion{Si}{4} profiles in the flare ribbon. \textit{k}-means clustering is an unsupervised machine learning algorithm that operates by iteratively partitioning a dataset into groups, based on their distance to a predefined quantity \textit{k} of means.  The method is advantageous in its ability to extract similar observations in a potentially large dataset with linear time complexity.  To begin the algorithm, the chosen means are arbitrary, and each observation is assigned to the nearest mean based on the least-squares Euclidean distance.  For each iteration thereafter, the means are defined as the centroid of the observations assigned to each cluster.  The algorithm is complete once the assigned distances to cluster centers stabilize.  This method is repeated 10~times with different initial cluster centers.  The most common output means are selected as the resulting clusters. A more mathematical description of applying \textit{k}-means clustering to the classification of solar spectral data is outlined in \citet{panos2018}. 

We apply the clustering analysis to the masked area of the southern ribbon highlighted in Figure~\ref{fig:IRIS_context}. This is the same region we average over for the velocity and intensity time series in Figure~\ref{fig:time_series}. All pixels within this mask are included in the clustering analysis, with the exception of 106 saturated pixels, which we exclude. In order to minimize the impacts of \ion{Si}{4} intensity and Doppler velocity on the outcome and instead focus on changes in line profile shapes, we normalize the spectra of each pixel by the spectral line amplitude, and normalize the line center to zero. These line profiles are fed into the clustering algorithm, which assumes each profile to be composed of 74 ``features,'' corresponding to the intensities $I_{\lambda}$ at wavelengths $\lambda$. The algorithm outputs representative \textit{typical} line profile shapes; each spectrum is associated with one of these clusters. These representative profiles are displayed as the colored profiles in Figure~\ref{fig:clustering} (bottom). The clusters in the figure are sorted from blue to dark red, based on their full width at 30\% maximum. We chose to order them via this definition of line width, instead of the more standard full width at half-maximum (FWHM), to account for some of the line profiles that split into two peaks below the half-maximum intensity. The top panel of Figure~\ref{fig:clustering} presents a colored spatial map with the locations of pixels within each clustering group.
We experimented with the number of clusters and empirically determined that $k = 40$ clusters sufficiently capture the diversity of line profiles, while minimizing redundancy in representative cluster profile shape. We note, however, that the exact number of clusters and cluster-center line profile shapes are not crucial to our analysis.

The K-means clustering map in Figure~\ref{fig:clustering} presents the simplest line profiles with blue to green colors, mostly representing spectra close to a Gaussian or asymmetrical-Gaussian shape. Contrastingly, the orange and red pixels represent more complex line profiles, such as those with multiple peaks or broadened, flat peaks. At the start of the flare, during 12:00--12:15~UT, almost all of the flare spectra are represented by these more complex line profiles. This time period corresponds to the most impulsive part of the flare, during the phase of strong HXR and Lyman-$\alpha$ emission, alongside the first stage of strong \ion{Si}{4} Doppler and intensity oscillations. Later into the flare, beyond 12:18~UT, the flare ribbon line profiles become less complex, represented in the clustering map by the blue and green pixels. However, although the majority of pixels during this later phase remain in these less complex clusters, concentrated regions of more complex line profiles (yellow to red in the map) are apparent within the 12:21--12:28~UT time window. In the strongest of these red-clustered pixels, the line profiles are in similar clusters to earlier pixels observed at flare onset, and thus possess similar line complexities to the pixels at the peak of the flare. These regions of red clusters correspond temporally to the sharp spikes in $>$100~km/s Doppler velocities displayed in Figure~\ref{fig:time_series}.

\section{Discussion}

In this work, we have analyzed \ion{Si}{4} 1402.77~{\AA} spectra of an X9-class solar flare from 2024 October 3rd. We find two distinct stages of rapid \ion{Si}{4} redshift velocities, each with different behaviors, suggesting that each stage of the downflows is facilitated by a different mechanism. Despite this difference, we find a constant oscillation period throughout the observations.

The first stage ($\sim$12:15--12:19~UT) of chromospheric downflows relates to the impulsive phase of the flare, coinciding with strong HXR and Lyman-$\alpha$ emission, with bursts in emission at these wavelengths coinciding with increases in \ion{Si}{4} redshifts. The HXR spectra also exhibit a strong non-thermal component during this time, consistent with significant particle acceleration. We measure \ion{Si}{4} redshifts up to 176~km/s during this impulsive phase. The line profiles observed during times of rapid redshifts are well-fit by a triple Gaussian profile, similar to those observed during a previous X-class event \citep[][who detect \ion{Si}{4} redshifts of 160~km/s]{Xu2023}, and consistent with previous interpretations that rapid flare ribbon redshifts can originate from the downwards component of the expansion of the chromosphere (chromospheric condensation), due to the rapid deposition of energy from waves or non-thermal particles originating from the reconnection site \citep[e.g.,][]{Ashfield2021,Kerr2022,Xu2023}.

The second stage of rapid \ion{Si}{4} downflows comes past the peak of the flare ($\sim$12:23--12:28~UT), and does not correlate to any of the more energetic signatures present with the earlier rapid downflows. Although notable 10--20~keV and 20--50~keV HXR emission is still present, HXR spectral analysis reveals that the remaining emission comes almost entirely from hot thermal emission, with only a very weak non-thermal component present (over two orders of magnitude weaker than during the first phase). From magnetic flux measurements, we find a lack of evidence for substantial ongoing reconnection at this time. The absence of signatures of particle acceleration or notable ongoing reconnection suggests that the \ion{Si}{4} downflows are less likely to relate to the flare energy release process, and is more in line with previous observations of flare-induced coronal rain, formed as thermal plasma cools from higher temperatures into the temperature sensitivity range of \ion{Si}{4} 1402.77~{\AA} emission. Located within the flare ribbons, the observed rapid downflows would therefore correspond to the base of the post-flare loops through which the coronal rain is present. The SJI 1330~{\AA} observations in the top panel of Figure~\ref{fig:IRIS_context} support this hypothesis, as the appearance of chromospheric-temperature flare loops corresponds to this second stage of \ion{Si}{4} oscillations. Furthermore, we see the IRIS slit intersecting the bright flare loop / coronal rain footpoint in the southern ribbon in the example SJI frame at 12:25:27~UT.
We note that our observed downflows (150--217~km/s) are faster than previous plane-of-sky observations of post-flare \ion{Si}{4} coronal rain downflows \citep[60~km/s in an X2.1-class flare and 100~km/s in an M-class flare in][respectively]{Lacatus2017,Yu2022}, although both of these measurements occurred within flares much less energetic than the X9-class event studied in this paper. To our knowledge, the detection of 217~km/s in \ion{Si}{4} is the fastest reported to date in a solar flare.
Previous spectroscopic observations of flare-induced coronal rain footpoints are also lower than what we observe in this event \citep[137~km/s in an X1.3-class flare][]{Song2025}. \citet{Pietrow2024} collected similar spectral measurements for the X13-class flare on 2017 September 6th (a stronger event than the flare presented in this study), but their instrument was unable to measure Doppler shifts greater than the observed 70~km/s. 
Previous work by \citet{Jing2016} found that brightenings associated with coronal rain are lower than the intensity of impulsive flare ribbons. With peak intensities two orders of magnitude fainter than the impulsive flare ribbons present ten minutes prior, our findings here are in agreement with this. Contrastingly, \citet{Jing2016} also found coronal rain footpoints to form behind the flare ribbon. This is to be expected in flare ribbons that are still propagating, as the coronal rain occurs within field lines that reconnected earlier than the field lines connected to the ribbon's leading edge. However, for the event presented in this work, Figure~\ref{fig:IRIS_context} reveals that the flare ribbons are no longer in motion by the time of the second stage of rapid downflows. In such a scenario, we would not expect the coronal rain footpoints to remain behind the flare ribbon, but to eventually ``catch-up'' with the now-stationary ribbon front (either before or after becoming visible in the \ion{Si}{4} line). This is consistent with our observations.

We note that the IRIS spectroscopic observations are collected from a sit-and-stare slit, probing only a limited region of the flare. In the first stage of \ion{Si}{4} oscillations, similar periods are also detected in full-disk HXR and Lyman-$\alpha$ emission. This makes it likely that the primary \ion{Si}{4} behavior is present across the wider flare region, and not just under the narrow slit. 
In the later stage of rapid \ion{Si}{4} oscillations, the IRIS slit covers the footpoint of a coronal loop. We therefore cannot say with certainty if similar Doppler and intensity oscillations are present elsewhere along the ribbons, but we can make the reasonable assumption that the behavior observed under the slit is unlikely to be radically different to behavior elsewhere along the flare arcade and ribbon structures. This scenario does, however, outline the future need for multi-slit spectroscopic observations of solar flares, such as that to be provided by the upcoming Multi-slit Solar Explorer \citep[MUSE;][]{DePontieu2020,Cheung2022} mission.
Figure~\ref{fig:time_series} does reveal finer, smaller-period oscillatory features in the \ion{Si}{4} time series, which may relate to more localized processes. Furthermore, we note that \citet{Li2025} present analysis of even shorter-period ($\sim$1~s) HXR oscillations for this flare.
Analyzing the origins of these features is beyond the scope of this work.

The fastest of the rapid \ion{Si}{4} redshifts are not constant, but oscillate as QPPs. Curiously, although the two stages of strong redshifts are likely produced by different plasma acceleration processes, the same oscillation period is consistent throughout both stages (albeit with different amplitudes and oscillation baselines). Given our interpretation that only one of the stages of oscillatory downflows is a result of impulsive magnetic reconnection, a constant period throughout both stages offers tantalizing insights into the origin of QPPs in solar flares. QPP models fall broadly into two camps -- signatures of bursty reconnection, or a result of MHD waves \citep[see reviews by][]{nakariakov2009,vandoorsselare2016,McLaughlin2018,Kupriyanova2020,Zimovets2021}. Because our second stage of fast redshift pulsations is not accompanied by any significant signatures of magnetic reconnection, this means that MHD oscillations are likely the cause of the QPPs in this event. Furthermore, many prime MHD wave candidates in solar flares have a period reliance on loop length. This includes popular MHD mechanisms like the global sausage mode \citep{Nakariakov2012,Tian2016,Nakariakov2018}. Because the oscillation period we observe remains close to constant throughout the flare evolution (as the flare loops grow), we can rule out MHD mechanisms with a period dependent on loop length.

Although we lack the diverse observational parameters needed to conduct a full comparison between proposed MHD processes for this event, we do note similarities between this flare and the observations presented in \citet{French2024b}. They analyzed IRIS coronal \ion{Fe}{21} observations of a flare fan/looptop, close to in phase with HXR sources at the ribbon footpoint. The crucial similarities between their study and the oscillations presented in this work are a common period of 50~s, persisting for over 12~minutes from flare onset. Due to the diverse multi-wavelength observations available for the event studied by \citet{French2024b}, they were able to determine the origin of the QPPs as the ``magnetic tuning fork'' mechanism, an oscillation in the flare looptops and fan region triggered by reconnection outflows \citep{Takasao2016,Reeves2020,Shibata2023}. Magnetic tuning fork oscillations do not require the reconnection outflows to be periodic, and the predicted period \citep[10--100~s;][]{Takasao2016} does not depend on loop length. In the magnetic tuning fork, HXR emission can oscillate due to betatron acceleration \citep{Brown1975}, as quasi-periodic changes in the cross-sectional area of magnetic loops allow quasi-periodic bursts of electrons to propagate down to the chromosphere, due to changes in the magnetic mirroring effect. At the start of the flare, this mechanism could create quasi-periodic energy deposition in the chromosphere and explain the observed 50~s periodicity of \ion{Si}{4} parameters, and their correlation to HXR and Lyman-$\alpha$ emission. Later in the flare, when the impulsive fast-reconnection is ended, the magnetic loops may continue to oscillate due to earlier (or weaker, ongoing) activation from reconnection outflows. As loop plasma cools to \ion{Si}{4} temperatures, and we see the footprint imprint of flare coronal rain in the flare ribbon, the loops would still be oscillating at the same period observed previously in the flare -- potentially explaining the similar-period QPPs present across both stages of the observed \ion{Si}{4} ribbon behavior. 

We also note further similarities between the results presented here, and recent work by \citet{Ashfieldetal2025}, who detected in-phase oscillations between IRIS \ion{Si}{4} Doppler velocity and ASO-S/HXI HXR emission, but with a period of $\sim$32~s. This QPP behavior was interpreted as evidence of periodic reconnection in the observed flare and resembles the first phase of \ion{Si}{4} Doppler oscillations presented in our work. For the M-class flare examined by \citet{Ashfieldetal2025}, however, there is no evidence of a second phase of strong oscillating \ion{Si}{4} Doppler velocity downflows with absent HXR emission. Whether the absence of this second phase of oscillations reflects differences in specific flare conditions, such as those that would generate coronal rain, or a larger departure from the QPP mechanism identified in this work, remains an open question for future investigation. 


Throughout all phases of the flare, the IRIS \ion{Si}{4} 1402.77~{\AA} line exhibits complex line profiles in regions of fast redshift velocities. Although these lines are unusual and complex, they broadly fit well with a triple-Gaussian fit. 
Typically, triple-Gaussian profiles are fitted on the assumption of an optically-thin expanding chromosphere/transition region, but the line profiles themselves have wide asymmetric widths, long tails, or small bumps. However, for the fastest redshift pixels observed in the flare, the triple-Gaussian shape is well resolved by three individual peaks. This profile shape is likely possible as rapid redshifts exceed the enhanced line widths for the \ion{Si}{4} emission. In some pixels, the line profiles even start to display behavior that may be consistent with absorption within the \ion{Si}{4} 1402.77~{\AA} line, which is typically assumed to be optically thin \citep[e.g.,][]{Brannon2015,Kerr2019b,Ashfield2022,Kerr2022,Ashfieldetal2025}. In-depth modeling efforts and a more detailed examination of line profiles, beyond the scope of this paper, are needed to confirm this. 

\section{Conclusions} \label{sec:concl}

We use spectroscopic \ion{Si}{4} and HXR analysis to conclude that rapid flare ribbon downflows can originate from multiple mechanisms, including both chromospheric condensations during the impulsive phase of the flare and flare-induced coronal rain at the base of post-flare loops. Although the plasma acceleration processes are different during these stages, common processes may trigger similar QPP oscillation periods during both. The magnetic tuning fork oscillation, and associated betatron emission, is a strong candidate for producing consistent QPP periods across both flare stages, but we lack the diversity of observation parameters to claim this definitively -- future missions observing in ultraviolet and HXRs are needed to distinguish between and constrain multiple competing explanations.
Finally, machine learning efforts reveal complex line profiles in regions of fast redshift velocities throughout the flare, identifying additional features of interest worthy of further investigation and modeling.

The analysis in this paper presents initial results from what is a rich and complex dataset from IRIS. There are many facets of the IRIS 2024 October 3rd X9-class flare data that we don't explore here, but could be of interest to the wider solar flare community. These include the tantalizing faster-period oscillations (which we smooth out in our analysis), signs of potential absorption in the \ion{Si}{4} 1402.77~{\AA} spectra, and further spectroscopic measurements of the \ion{C}{2} 1336~{\AA} and \ion{Mg}{2}~k 2796~{\AA} lines. There are also several hours of pre-flare data available for this event. This IRIS dataset will likely be analyzed by the solar flare community for years to come. 

\begin{acknowledgments}
IRIS is a NASA Small Explorer mission developed and operated by LMSAL with mission operations executed at NASA Ames Research Center and major contributions to downlink communications funded by ESA and the Norwegian Space Centre. LYRA is a project of the Centre Spatial de Liege, the Physikalisch-Meteorologisches Observatorium Davos and the Royal Observatory of Belgium funded by the Belgian Federal Science Policy Office (BELSPO) and by the Swiss Bundesamt für Bildung und Wissenschaft
R.J.F. thanks support from NASA HGI award 80NSSC25K7927. C.A.T. and M.C.A. acknowledge the support of the the DKIST Ambassadors program, administered by the National Solar Observatory and AURA, Inc. W.A. acknowledges support from NASA HGI grant 80NSSC24K0451. M.D.K. acknowledges support by NASA ECIP award 80NSSC19K0910 and NSF CAREER award SPVKK1RC2MZ3. M.D. acknowledges support from BELSPO in the framework of the ESA-PRODEX program, PEA 4000145189. A.C. was supported in part by NASA HFORT awards 80NSSC22M0111 and 80NSSC22M0098, and by NASA LCAS award 80NSSC24M0030.
\end{acknowledgments}

\bibliography{bibliography}{}
\bibliographystyle{aasjournal}



\end{document}